\documentclass[lettersize,journal]{IEEEtran}
\usepackage{amsmath,amssymb,amsfonts}
\usepackage{algorithmic}
\usepackage{algorithm}
\floatname{algorithm}{Method}

\usepackage{array}
\usepackage[caption=false,font=normalsize,labelfont=small,textfont=sf]{subfig}
\usepackage{textcomp}
\usepackage{stfloats}
\usepackage{url}
\usepackage{verbatim}
\usepackage{graphicx}
\usepackage{cite}

\usepackage{makecell}
\usepackage{booktabs}
\usepackage{multirow}
\usepackage{tabularx}
\usepackage{pifont}
\newcommand{\cmark}{\ding{51}}
\newcommand{\xmark}{\ding{55}}
\usepackage{siunitx}

\DeclareRobustCommand*{\IEEEauthorrefmark}[1]{%
  \textsuperscript{#1}%
}
\begin{document}
\title{Co‑Fabric: Breaking Host‑Domain Boundaries for Unified xPU Interconnection}
\author{
\IEEEauthorblockN{Zhen Peng\IEEEauthorrefmark{1},
Jiaming Huang\IEEEauthorrefmark{1},
Chaofan Chen\textsuperscript{*}\IEEEauthorrefmark{1},
Zhao Zhang\textsuperscript{*}\IEEEauthorrefmark{1},
An Wu\IEEEauthorrefmark{1},
Baoyang Liu\IEEEauthorrefmark{2},
Xinglong Wang\IEEEauthorrefmark{2},
Tanlong Ci\IEEEauthorrefmark{2},
Jinfeng Li\IEEEauthorrefmark{1},
Xueke Duan\IEEEauthorrefmark{1},
Hao Wang\IEEEauthorrefmark{2},
Xi Chen\IEEEauthorrefmark{2},
Shunshun Zhang\IEEEauthorrefmark{2},
Zhiyuan Su\IEEEauthorrefmark{2},
Zhu Cao\IEEEauthorrefmark{2},
Zhichong Dou\IEEEauthorrefmark{2},
Shaohua Wu\IEEEauthorrefmark{1},
Lu Jing\IEEEauthorrefmark{1},
Yue Yuan\IEEEauthorrefmark{1}}

\IEEEauthorblockA{\IEEEauthorrefmark{1}IEIT SYSTEMS Co., Ltd., Beijing, China}
\IEEEauthorblockA{\IEEEauthorrefmark{2}IEIT SYSTEMS Co., Ltd., JiNan, China}

\thanks{*Corresponding authors. E‑mail: chenchaofan@ieisystem.com, zhangzhao06@ieisystem.com.}
}

\markboth{PENG \textit{et al.}: CO‑FABRIC: BREAKING HOST‑DOMAIN BOUNDARIES FOR UNIFIED XPU INTERCONNECTION}%
{PENG \textit{et al.}: CO‑FABRIC: BREAKING HOST‑DOMAIN BOUNDARIES FOR UNIFIED XPU INTERCONNECTION}

\maketitle

\begin{abstract}
Large-model parameters have grown beyond the capacity of a single xPU, dispersing across multiple xPUs spanning distinct host domains, where xPU-to-xPU communication dominates overall system efficiency. Existing scale-up interconnect remains inadequate: network-based solutions built on Ethernet—such as RoCE (RDMA over Converged Ethernet)—introduce specific message-semantics and protocol-stack characteristics, and rely on fragmented per-host addressing, while conventional host-based fabrics are confined to a single host domain and lack cross-host unified addressing. This paper presents Co-Fabric, a bus-based interconnect that, unlike conventional bus designs, breaks host-domain boundaries to deliver unified xPU interconnection for scale-up superpods. Co-Fabric makes three contributions: a streamlined four-layer protocol stack achieving nanosecond-scale processing latency with native reliability; a cross-domain scaling and P2P mechanism that routes using port identifiers embedded in the packet header; and a unified address space built on shadow-device auto-enumeration. On a 64-xPU 3D-Mesh system, Co-Fabric cuts inter-node communication latency by over 50\% and improves bandwidth by 2–5× over RoCE, accelerating DeepSeek R1 inference by 30\%-80\%. Moreover, since its streamlined four-layer protocol stack and higher data-communication efficiency reduce protocol and processing overhead relative to the Ethernet-based RoCE stack, Co-Fabric cuts the cost and power of the interconnect itself by up to 80\% and 5\%, respectively. These results demonstrate Co-Fabric's advantage for AI computing centers.

\end{abstract}


\section{Introduction}
\label{sec:introduction}

Scaling laws dictate that stronger model capability requires larger model scale, and as model parameters grow to trillions, a single xPU or host can no longer accommodate the entire model replica \cite{kaplan2020scaling,hoffmann2022training}. Model parameters must be distributed across multiple xPUs and nodes, making communication performance the decisive factor in overall system efficiency. The industry has converged on the scale-up superpod \cite{qian2025research} as the architectural answer: interconnecting tens to hundreds of xPUs at high speed so they behave as a single giant accelerator. At this scale, the defining challenge is no longer the xPU itself but the interconnect protocol---the set of rules governing data movement across devices, host domains, and physical links. How to design an interconnect protocol that delivers a unified address space and low-latency hardware routing across host domains is the central question this paper addresses.

Industry efforts to build scale-up interconnect split into two distinct design philosophies. The bus-based camp treats xPU interconnect as a bus that extends the compute domain: it builds a dedicated compute bus domain in which every device participates directly in data exchange, delivering high routing efficiency, low latency, and native memory semantics. This paper belongs to this camp, and builds on its strengths while addressing its limitations in cross-domain reach. The network-based camp, in contrast, treats xPUs as nodes on a network: it builds on the mature Ethernet ecosystem, reusing rapid SerDes evolution and established flow control to achieve fast commercialization and broad openness. This route, however, is a compromise—it starts from the existing networking vantage point and adapts it to interconnect, sacrificing the latency and memory-semantics advantages of a bus domain and inheriting the encapsulation and software-stack overhead of network protocols.

In summary, the network-based approach pays for the convenience of reuse with higher latency and coarser memory semantics. The bus-based approach, while architecturally superior, remains largely confined to xPU-to-xPU fabrics with limited cross-domain reach and no native cross-host unified memory access. Neither camp currently delivers cross‑domain unified addressing, native memory semantics, and hardware‑level low‑latency routing across host domains. Recognizing the architectural strengths of the bus-based philosophy while addressing its cross-domain limitations, this paper proposes a new bus-based interconnect architecture that extends these capabilities across host boundaries.

Drawing from the above analysis of industry practice, we identify six design principles for scale-up interconnect protocols, grounded in the demands of contemporary workloads such as agent inference and MoE-based training \cite{pan2025fsmoe}.

\begin{itemize}
    \item \textbf{Streamlined protocol layering.} 
    Multi-layer network protocol stacks (TCP/IP or OSI models) introduce processing latency at every encapsulation boundary \cite{hunt2002tcp, alani2014guide}. For scale-up scenarios where communication patterns are predictable and confined within a superpod, such generality is unnecessary. A protocol designed for scale-up should condense its stack to the minimum layers required, eliminating dedicated transport layers and traditional multi-hop routing overhead while integrating essential routing and reliability functions directly into its core layers rather than delegating them to separate transport and network abstractions.

    \item \textbf{Memory consistency support.} For superpod deployments, memory consistency\cite{sorin2011primer} offers well‑defined ordering semantics for remote memory reads and writes. This capability enables transparent programming, while circumventing the substantial complexity of enforcing full cache coherence across all xPUs.

    \item \textbf{Native memory semantics.} Load/store instructions are the most natural primitive for xPU programmers \cite{sharma2024pci}. A protocol that natively supports memory semantics allows applications to access remote memory as if it were local, eliminating the impedance mismatch of message-passing APIs and the associated data copying overhead.

    \item \textbf{Ultra-low communication latency.} Workloads such as MoE All-to-All and agent inference involve frequent, small-granularity exchanges \cite{liu2026ubep}. Here, the intrinsic protocol latency---not bandwidth---becomes the dominant factor. Every nanosecond of protocol overhead is magnified by the sheer frequency of these operations.

    \item \textbf{Sufficient communication bandwidth.} While bandwidth need not be extreme, it must match the aggregate compute throughput of the xPU pool. Insufficient bandwidth creates a starvation scenario where xPUs idle waiting for data.

    \item \textbf{High reliability.} At scale-up deployments, the probability of link errors increases. The protocol must natively support flow control and link-level retransmission \cite{liu2026evaluating}, rather than relying on upper-layer retry mechanisms or network-level congestion control, which introduce unacceptable latency jitter.
\end{itemize}

To the best of our knowledge, no existing interconnect protocol satisfies all six design principles simultaneously. Existing approaches are typically forced to choose between low-latency memory semantics within a limited domain and cross-domain connectivity with higher protocol overhead, leaving a fundamental gap for unified scale-up interconnects. This gap motivates the design of Co-Fabric, an interconnect protocol purpose-built for scale-up superpods.

The main contributions of this paper are as follows:

(1) We propose Co-Fabric, an interconnect-centric architecture and protocol that unifies computing, storage, network, and management resources into a single interconnect domain. Its streamlined four-layer protocol stack achieves nanosecond-level latency with native reliability---credit-based flow control and link-level retransmission---while topology disaggregation enables elastic scaling from full-mesh direct connection to thousands of xPUs, addressing the fragmentation between intra-domain xPU fabrics and inter-domain network fabrics.

(2) We design a cross-domain scaling and peer-to-peer (P2P) communication mechanism that enables direct xPU-to-xPU data transfer across host boundaries without traversing the host CPU or network protocol stack. Through port-ID routing implemented entirely in switch hardware, Co-Fabric achieves native cross-domain addressing with routing overhead confined to nanoseconds, eliminating the explicit memory registration and message-passing semantics inherent in RDMA-based solutions.

(3) We introduce a global unified address space mechanism based on shadow device auto‑enumeration. Each host discovers all xPUs in the superpod as locally attached devices through a flat unified address space, allowing developers to write distributed multi‑xPU code using standard single‑node programming models without topology awareness or explicit data migration logic.

The remainder of this paper is organized as follows. Section~\ref{sec:hardware-partitioned-system} reviews the background and motivation for scale-up interconnect design. Section~\ref{sec:architecture} details the Co-Fabric architecture, covering its unified interconnect design, streamlined protocol layering, cross-domain scaling and P2P mechanism, global unified address space, and low-latency/high-reliability techniques. Section~\ref{sec:prototype} presents the Co-Fabric AI Scaling system. Section~\ref{sec:analysis} presents experimental results and analysis. Section~\ref{sec:conclusion}  concludes this paper.

\section{Background  \&   Motivation}
\label{sec:hardware-partitioned-system}

\subsection{The Memory Wall and the Data Movement Bottleneck}

Modern AI accelerators exemplify a systemic imbalance: compute capacity has consistently outpaced memory bandwidth across xPU generations. While single-chip floating-point throughput has grown by orders of magnitude---driven by expanding core counts, higher clock frequencies, and specialized tensor units---memory bandwidth has improved at a far more modest rate. This widening compute-to-bandwidth ratio means that xPUs can rarely sustain peak utilization; they spend an increasing fraction of time stalled, waiting for data to arrive from memory. This phenomenon, widely recognized as the \textit{memory wall}, has become the primary performance limiter in large-scale AI systems.

The problem extends beyond individual xPU memory. At the superpod level, the overall system is similarly over-provisioned in compute relative to storage and data-access capacity. Model parameters, optimizer states, activation checkpoints, and KV caches must be staged from host memory or storage devices into xPU HBM before computation can proceed. When the data supply chain---spanning persistent storage, host DRAM, and xPU HBM---cannot keep pace with compute demand, the memory wall manifests at the system level, not just within a single chip.

This bottleneck is further compounded by the communication demands of distributed execution. Large language model training and inference rely on multiple parallelism strategies---data parallelism, tensor parallelism, pipeline parallelism, and expert parallelism---each of which requires frequent inter-xPU data exchange. AllReduce operations synchronize gradients across replicas; tensor-parallel collectives exchange intermediate activations at every layer; pipeline stages pass activations forward and backward; and Mixture-of-Experts (MoE) models dispatch tokens to remote experts via All-to-All patterns. When inter-xPU communication is slow, xPUs stall waiting for data that resides on a peer or a distant host, effectively reducing the rate at which they can access memory. In other words, inefficient communication does not merely add latency---it directly exacerbates the memory wall by throttling the effective data supply to each xPU.

This analysis leads to a central observation: the performance of a superpod system is governed not by raw compute capacity, but by the efficiency of \textit{data movement}---the rate at which data can be transported across the full path from storage to host memory to xPU HBM and across xPUs. Data movement efficiency is, in turn, determined by the interconnect architecture. Only by optimizing interconnect design---reducing protocol overhead, enabling direct peer-to-peer access, and unifying the address space across host domains---can the memory access bottleneck be mitigated and overall system performance improved.

\subsection{Limitations of Existing Scale-Up Interconnect Approaches}

The industry has pursued two distinct routes to scale-up interconnect design, each addressing the data movement challenge from a different architectural foundation.

The first route builds a dedicated compute-domain bus interconnect, exemplified by NVIDIA's NVLink. By providing high-bandwidth, low-latency links between GPUs, NVLink enables rapid peer-to-peer data transfer that significantly improves data movement efficiency within the GPU domain. This bus-based design delivers the low-latency, memory-semantic communication that compute-intensive workloads demand. However, NVLink targets a GPU‑focused interconnect domain. Data that must traverse the boundary between the xPU fabric and the rest of the system still incurs the overhead of host-mediated transfers and protocol translation.

The second route builds a network-based interconnect using protocols such as RoCE (RDMA over Converged Ethernet) \cite{maniotis2025roce}. Compared to traditional Ethernet, RoCE offers advantages including kernel bypass, zero-copy data transfer, and hardware-offloaded transport, which collectively reduce CPU overhead and improve latency. Nevertheless, RoCE remains fundamentally a network-based architecture. Its communication model relies on message semantics: remote memory access requires explicit registration of memory regions, exchange of remote keys, and management of queue pairs. The data path traverses a multi-layer network protocol stack, and each host maintains an independent address space with no native cross-domain unified addressing. These architectural constraints impose cumulative overhead on every data movement operation, resulting in data movement efficiency that falls short of bus-based approaches---particularly for the fine-grained, high-frequency communication patterns characteristic of large-model training.

Neither route fully resolves the data movement bottleneck. The bus-based paradigm has demonstrated superior data movement efficiency within the xPU domain, yet existing implementations confine it to xPUs---storage, network, and management resources remain outside the unified fabric. The network-based approach, while enabling broader connectivity, incurs protocol overhead and fragmented addressing that limit its data movement efficiency. This gap motivates Co-Fabric, a bus-based interconnect architecture---interconnect-centric by design---that extends the compute-domain bus paradigm beyond the xPU domain to encompass all system resources (computing, storage, network, and management) within a single interconnect fabric. By providing cross-host-domain unified addressing, native P2P communication with load/store semantics, and a streamlined protocol stack, Co-Fabric aims to establish a unified data movement plane that breaks the host-domain boundary and maximizes end-to-end data movement efficiency across the entire superpod.

\section{Co-Fabric Architecture}
\label{sec:architecture}
The prefix ``Co-'' in Co-Fabric encapsulates three design intentions that shape the interconnect architecture.
\textit{Coalescent}: heterogeneous system resources---computing, storage, network, and management---are merged into one resource pool, dissolving conventional boundaries among device classes.
\textit{Coequal}: all connected devices hold peer status, with no device type designated as the topological root, enabling symmetric, host-independent communication.
\textit{Coherent}: the fabric maintains logical unity across host operating system domains; every host observes a consistent system view---a unified flat address space, identical communication semantics, and transparent device enumeration.
Together, these three intentions define an interconnect that is functionally whole rather than merely connected.
The following subsections detail how these intentions are realized through Co‑Fabric's interconnect architecture design, streamlined protocol layering, cross‑domain scaling mechanisms, and global address space unification.

\subsection{Interconnect Architecture Design}
\begin{figure*}[htbp]
\centering
\includegraphics[width=0.78\textwidth]{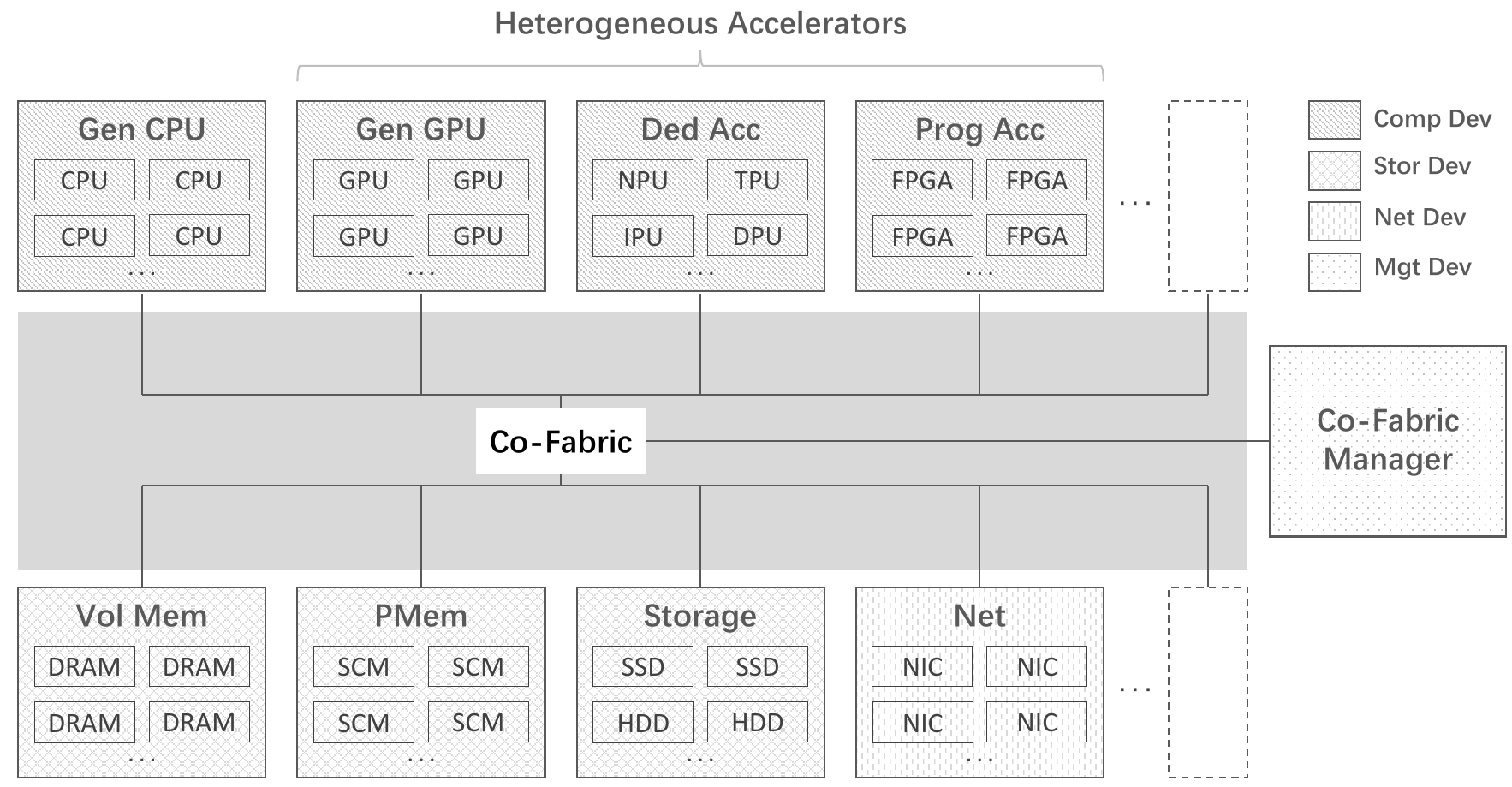}
\caption{Overview of Co-Fabric architecture with unified integration of heterogeneous computing, storage, network and management resources.}
\label{fig:3.1}
\end{figure*}

Co-Fabric proposes a bus-based AI superpod system architecture---interconnect-centric by design. Fig.~\ref{fig:3.1} illustrates its overall design. Taking the Co-Fabric interconnect as the unified backbone of the entire system, this architecture integrates computing resources (general-purpose CPUs, general-purpose GPUs, dedicated accelerators, and programmable accelerators), storage resources (volatile memory, persistent memory, and block storage devices), network resources (scale-out interfaces such as Ethernet), and management resources (management controllers including BMCs). The Co‑Fabric Manager is in charge of system‑wide topology management, system initialization, runtime configuration, and device‑level administration for the Co‑Fabric system.

Unlike conventional hierarchical interconnect architectures rooted in hosts or xPUs, Co-Fabric does not designate any single device type as the topological root; all connected devices hold equal status at the interconnect layer. This bus-based architectural design enables two critical capabilities: global unified address space mapping across operating system domains and native P2P cross-domain communication. Furthermore, data transmission over Co-Fabric links bypasses multi-layer encapsulation inherent to traditional network protocol stacks, laying a foundational architectural basis for ultra-low-latency communication.

Table~\ref{tab:arch_compare} compares Co-Fabric with RDMA-based interconnect solutions across key technical dimensions.

\begin{table*}[htbp]
\centering
\caption{Comparison of mainstream interconnect schemes}
\label{tab:arch_compare}
\begin{tabular}{p{2.2cm} p{3.7cm} p{3.7cm}}
\toprule
\textbf{Dimension} & \textbf{RDMA (RoCE/IB)} & \textbf{Co-Fabric} \\
\midrule
Architectural Core & network-based  & bus-based \\
\addlinespace[0.3em]  
Cross-OS-Domain Unified Address Space & \xmark \ Each host maintains independent address space & \cmark \ Cross-domain global unified addressing\\
\addlinespace[0.3em]  
Cross-Domain P2P Communication & \cmark \ Supported, based on message semantics, requires explicit memory registration & \cmark \ Native cross-domain P2P, load/store semantics, bypassing network \\
\addlinespace[0.3em]  
Interconnect Model & Host-independent, connected to network via NIC & Peer connectivity across multiple hosts and heterogeneous devices; unified integration of four categories of system resources \\
\addlinespace[0.3em]  
Scaling Paradigm & Native Scale-Out & Elastic scale-up, expandable to thousand-xPU clusters \\
\bottomrule
\end{tabular}
\end{table*}

RDMA (such as RoCE/InfiniBand) adopts a network-based architectural core: it supports cross-domain P2P communication through message semantics, but each host maintains an independent address space. Remote memory access requires explicit registration, and the data path traverses the network protocol stack.

Co-Fabric's interconnect-centric design eliminates the aforementioned limitations of RDMA. It consolidates all four categories of system resources into one homogeneous interconnect domain, with native architectural support for cross-OS-domain unified address spaces and P2P communication, thereby eliminating reliance on auxiliary external networks like Ethernet.

From a topological perspective, the Co-Fabric protocol stack is decoupled from underlying physical topologies. For small-scale training workloads with stringent latency requirements, xPUs can be wired in a full-mesh direct topology to eliminate forwarding latency induced by intermediate switching nodes. As deployment scale expands, Co-Fabric Switches enable elastic scaling from dozens of xPUs up to thousand-card clusters. Regardless of the underlying physical topology, the Co-Fabric protocol exposes a uniform link abstraction and consistent communication semantics to upper-layer software, which remains agnostic to topological variations. This topology-agnostic design grants Co-Fabric the flexibility to adapt to diverse deployment scenarios. At the protocol layer, a streamlined layered framework further minimizes data processing latency.

\subsection{Streamlined Four‑Layer Protocol Stack}

Co-Fabric implements a compact four-layer protocol stack, consisting of the Media Layer (ML), Link Layer (LL), Fabric Layer (FL), and Semantic Layer (SL) from bottom to top, as visualized in Fig.~\ref{fig:3.2}. The four layers feature well-defined functional boundaries and minimalist inter-layer interfaces, purpose-built to satisfy the ultra-low processing latency demands of scale-up interconnect scenarios. Meanwhile, in terms of protocol design, Co-Fabric considers compatibility with PCIe physical links to support heterogeneous computing resources including mainstream CPUs and xPUs commercially available. The responsibilities of each layer are elaborated as follows:

\begin{figure}[htbp]
    \centering    \includegraphics[width=0.40\textwidth]{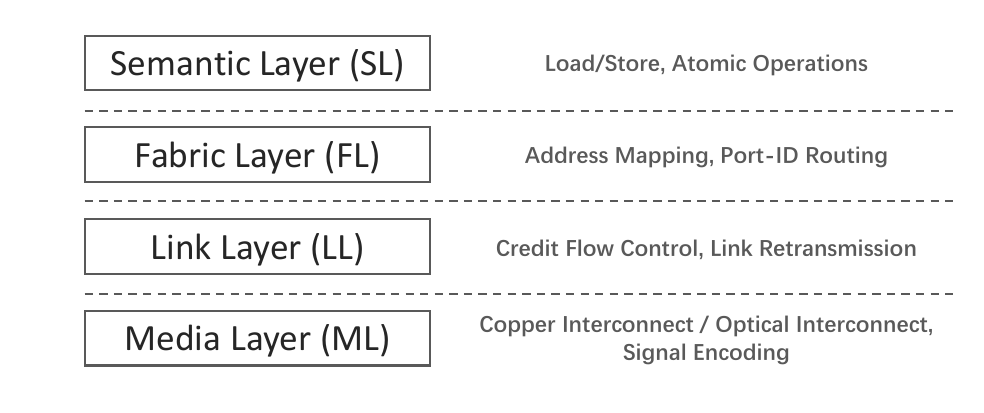}
    \caption{Four-layer streamlined protocol architecture of Co-Fabric.}
    \label{fig:3.2}
\end{figure}

\subsubsection{Media Layer (ML)}
This layer governs physical bitstream transmission and signal encoding/decoding. It abstracts heterogeneous physical media (copper and optical interconnects alike) from upper layers, insulating all upper layers from hardware-specific physical medium implementations.

\subsubsection{Link Layer (LL)}
The Link Layer guarantees reliable data delivery between adjacent nodes. Distinct from conventional network stacks that rely on end-to-end congestion control, Co-Fabric’s Link Layer leverages credit-based flow control to enable lossless transmission at the link level. It additionally supports link-level retransmission, localizing fault recovery within individual single-hop links rather than delegating reliability recovery to upper-layer end-to-end protocols, which caps latency overhead introduced by reliability mechanisms within a single hop.

\subsubsection{Fabric Layer (FL)}
The Fabric Layer manages global address space mapping and port-ID routing across the entire interconnect domain. It assigns globally unique port identifiers to every port in the switching fabric and maintains address-to-port mapping rules that translate OS-level device addresses into switch-internal routing decisions. The Fabric Layer also facilitates device discovery during system initialization, enabling automatic topology detection and routing table generation. Detailed routing mechanisms are elaborated in Section~\ref{subsec:p2p}.

\subsubsection{Semantic Layer (SL)}
The Semantic Layer defines Co-Fabric's core operational paradigm: native support for load/store memory semantics and atomic operations, which differentiates it from traditional network stacks built upon message-passing primitives. It translates remote memory access requests into protocol-level rail-onlys and exposes a transparent, streamlined memory-access interface to upper software stacks.

The ``streamlined'' nature of the Co-Fabric protocol design manifests in three key aspects:
\begin{enumerate}
    \item \textbf{Minimal Layer Count}
    Conventional general-purpose wide-area network stacks adopt four to seven layers (e.g., the four-layer TCP/IP model or seven-layer OSI model), incorporating dedicated transport layers for end-to-end reliability and network layers for multi-hop routing. Within scale-up interconnect environments, however, the Link Layer natively provides reliable transmission and flow control, while the Fabric Layer directly handles unified address routing---rendering dedicated transport and network layers redundant. Co-Fabric condenses the protocol stack into four layers, each implementing only functionality essential to scale-up interconnect workloads.

    \item \textbf{Simplified Communication Semantics}
    Top-layer interfaces of traditional network stacks expose sockets or message queues, forcing upper software to undergo repeated protocol encapsulation and data copying. By contrast, the Co-Fabric Semantic Layer adopts load/store as its fundamental primitive. Remote memory accesses initiated by xPUs are directly translated into protocol transactions at the semantic layer, yielding the shortest possible execution code path.

    \item \textbf{Compact Processing Pipeline}
    Data traverses the Co‑Fabric protocol stack along a unidirectional pipeline: Semantic Layer $\rightarrow$ Fabric Layer $\rightarrow$ Link Layer $\rightarrow$ Media Layer. No software‑based intermediate buffering or cross‑layer intervention occurs throughout transmission, restricting processing latency to the sub‑microsecond‑to‑microsecond range. This characteristic aligns perfectly with the requirements of high‑frequency, fine‑grained communication prevalent in scale‑up training workloads---such as all‑to‑all collective operations for mixture‑of‑experts models---where minimal inherent latency is critical.
\end{enumerate}

\subsection{Cross-Domain Scaling and ID-oriented Forwarding}
\label{subsec:p2p}

Scaling a superpod across multiple host domains and racks requires low‑latency, high‑bandwidth cross‑domain xPU communication. Network‑based solutions can cross domains, but they incur millisecond‑scale overhead from multi‑layer encapsulation and host‑side relay. Bus‑based solutions, in contrast, use hierarchical configuration‑space addressing confined to a single host domain and cannot cross domains at all. Our protocol is designed around two core principles to address this gap. First, minimalism: the protocol stack natively implements only the capabilities essential for cross‑domain communication, without adopting the transport and routing abstractions found in general‑purpose network stacks. Second, header‑embedded routing: all routing information is embedded in the packet header, so routing decisions complete automatically as the packet flows, without real‑time control‑plane intervention.

\begin{figure}[htbp]
\centering
\includegraphics[width=0.48\textwidth]{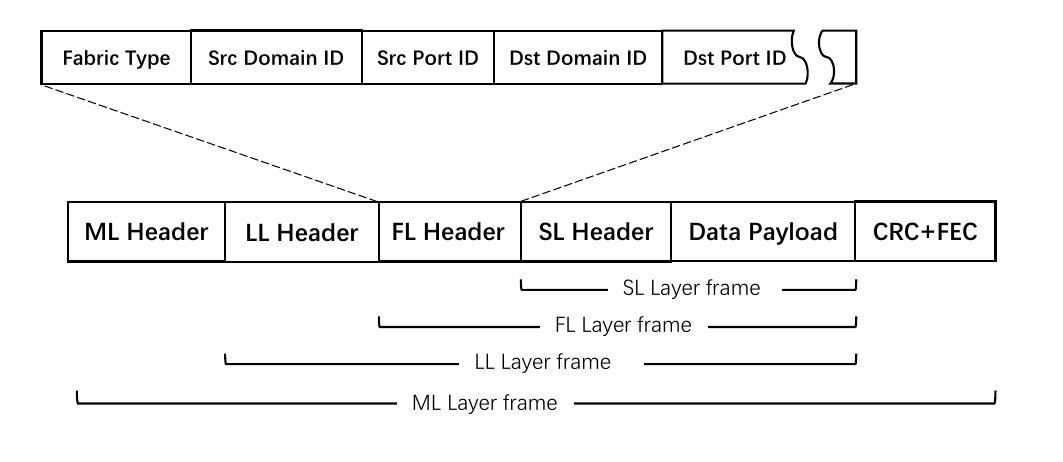}
\caption{Co-Fabric protocol stack.}
\label{fig:FL_protocal}
\end{figure}

These two principles are embodied in the Fabric Layer (FL) of the Co-Fabric protocol stack, as shown in Fig.~\ref{fig:FL_protocal}. The FL sits between the Link Layer and the Semantic Layer and handles cross-domain addressing and packet forwarding. The FL header carries the Fabric Type, source/destination Domain ID, and source/destination Port ID. Fabric Type indicates the transmission mode. Domain ID identifies the domain of the packet's source and destination; a domain corresponds to one OS's address space, so multiple OSes form multiple domains. Port ID identifies a specific port within a domain, covering both xPU ports and cascade ports between Switches. The FL header also carries other fields reserved for extension, which we do not detail here.

This design requires every port in the fabric to be assigned a globally unique identifier at initialization, along with routing-table configuration on all Switches. Firmware completes this process at power-on. Each Switch first discovers its downstream xPUs and peer Switches through topology self-discovery. Firmware then assigns globally unique Port IDs to all xPU ports and cascade ports in traversal order. Each Switch then installs its forwarding table and finally learns the domain ID of its ingress side. After this, a Switch keeps only one forwarding table and forwards a packet out the port indicated by its Port ID. Because all ports share globally unique IDs, routing across Switches needs no distributed negotiation, and the forwarding table stays stable once established.

With the FL header and unified numbering in place, a cross-domain read/write proceeds as follows. The source translates the target address into a port identifier at the Semantic Layer and fills the destination Domain ID and Port ID into the FL header, along with its own source Domain ID and Port ID. Address resolution happens at the endpoint; the FL header carries only IDs. Once the packet enters the fabric, each Switch parses only the FL header and forwards the packet out the port indicated by the destination Port ID. If the destination port is local, the Switch delivers it to the corresponding xPU port; otherwise the packet goes through a cascade port to the next Switch, repeating until it reaches the destination domain. Throughout forwarding, a Switch parses no address and never decapsulates the packet to the Semantic Layer; the Port ID alone suffices for routing. The destination xPU receives the packet and completes load/store semantics at the Semantic Layer. Response packets follow the reverse path: each Switch records the request source when forwarding, so the response returns along the original route. The path bypasses the host CPU and the network stack, and routing overhead is handled in parallel hardware at nanosecond scale.

The Fabric Type field lets a single packet format express multiple transmission modes. It supports unicast P2P, multicast, broadcast, and in-network computation (INC) \cite{jiang2026switch}, all executed in hardware with no software involvement in mode switching. Table~\ref{tab:fabric_type} lists the Fabric Type encodings and their forwarding behaviors.

\begin{table*}[htbp]  
\centering  
\caption{Fabric Type encodings and forwarding behaviors}  
\label{tab:fabric_type}  
\begin{tabular}{cll}  
\toprule  
\textbf{Encoding} & \textbf{Type} & \textbf{Forwarding Behavior} \\
\midrule  
000 & Unicast & Point-to-point pass-through, the default for most end-to-end traffic \\
001 & Multicast & One-to-many hardware replication to a member port list by group ID \\
010 & Broadcast & Floods all ports within a domain, for low-frequency control/ops traffic \\
011 & INC-ADD & Element-wise addition along the path, result forwarded to destination \\
100 & INC-MIN & Element-wise minimum along the path, result forwarded to destination \\
101 & INC-MAX & Element-wise maximum along the path, result forwarded to destination \\
110 & Reserved & Reserved for future operators or routing modes \\
111 & Reserved & Reserved for future operators or routing modes \\
\bottomrule  
\end{tabular}  
\end{table*}

Unicast is the default mode: a packet follows a single path to its destination Port ID, serving cross-domain P2P communication and remote reads/writes. In multicast mode, the destination Domain ID and Port ID positions in the FL header are reused as a multicast group ID; Switches along the path replicate the packet to the member port list in hardware, delivering one transaction to many destinations atomically---well suited to weight broadcast and KV-cache replication. Broadcast mode floods the packet to all ports in the destination domain, used only for low-frequency control and operations tasks such as configuration push and status query.

In-network computation offloads reduction onto the forwarding path. INC-ADD, INC-MIN, and INC-MAX correspond to element-wise addition, minimum, and maximum. A packet carries the operator type and initial value; INC-capable Switches apply the corresponding element-wise aggregation to the payload as they forward, and the aggregated result continues toward the destination. This completes a cross-domain reduction in a single transaction that would otherwise require many round trips, which is critical for collectives such as AllReduce.

Embedding the Port ID in the packet also makes routing decisions transparent to the header format, which enables linear scaling. The forwarding table size grows linearly with the number of ports in the domain, while the FL header width stays fixed regardless of scale. Adding Switch levels and extending forwarding tables scales a superpod from 16 xPUs to 32, 64, 128, 256, or more. When a new xPU joins, the Switch discovers it, assigns a unique Port ID, and installs the corresponding forwarding entry---no change to the header format or host software stack is needed.

\subsection{Global Unified Memory Addressing Mechanism}

\begin{figure}[htbp]
\centering
\includegraphics[width=0.48\textwidth]{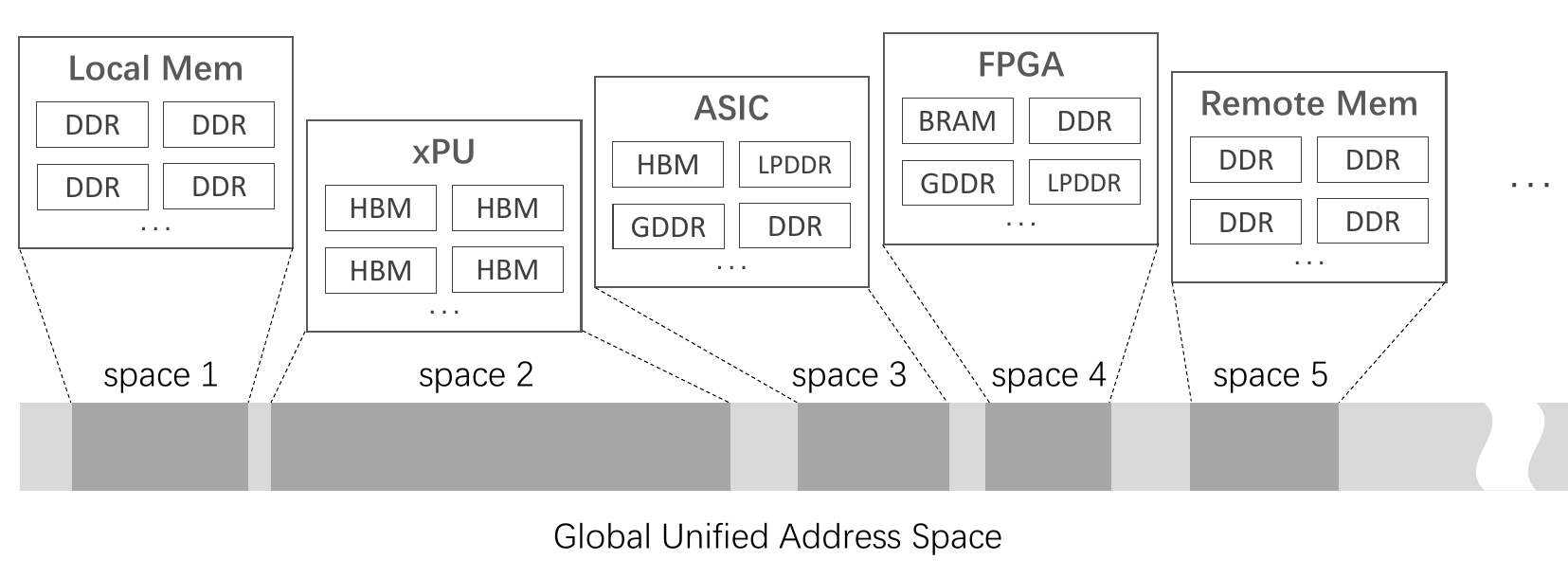}
\caption{Illustration of Unified Memory Address Space with Co‑fabric Interconnect.}
\label{fig:unified}
\end{figure}

\begin{figure}[htbp]
\centering
\includegraphics[width=0.48\textwidth]{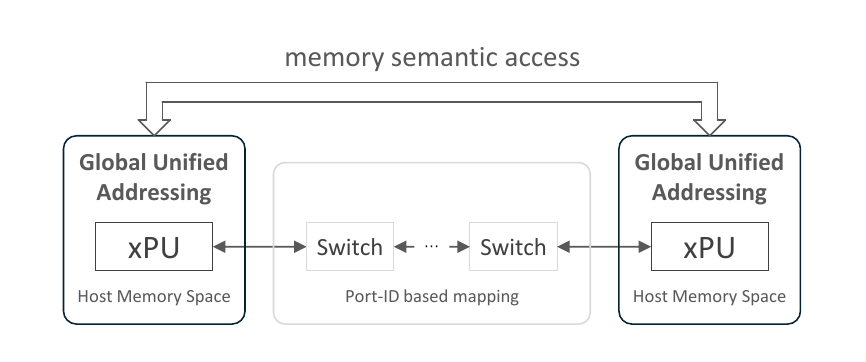}
\caption{Global unified addressing from a host-centric perspective: memory-semantic access to all xPUs, with Port-ID-based routing through Switch components.}
\label{fig:global-addressing}
\end{figure}

Co‑Fabric unifies the address spaces of all memory‑attached devices across the whole system, as shown in Fig.~\ref{fig:unified}. All nodes within this interconnect domain are mapped into one single large global unified address space, which partitions distinct non‑overlapping address ranges for heterogeneous memory resources, including host local memory, xPU on‑package HBM, ASIC‑attached multi‑type memory, FPGA‑integrated on‑chip and off‑chip memory, as well as remote memory from distant nodes. From a host‑centric perspective, every host sees all xPUs---both local and remote, intra‑domain and cross‑domain---uniformly mapped into its own host memory space, enabling memory‑semantic access to the memory of any xPU in the superpod through local load/store instructions, without distinguishing whether the target resides within the same host domain or beyond. As depicted in Fig.~\ref{fig:global-addressing}, all transactions follow the Co‑Fabric protocol for packetization and routing: access paths that traverse Switch components are forwarded via Port‑ID‑based mapping, while direct paths that do not pass through a Switch are routed by the global unified address itself.

Global unified addressing thus lies at the core of the Co-Fabric protocol, serving as the foundation for uniform, memory-semantic access across the entire superpod. Taking xPU as an example, we elaborate on the implementation of this unified address space: in a Co-Fabric superpod, xPUs spread across different host domains appear as local devices on the host's own interconnect bus, and developers assign multi-xPU tasks using standard single-node CUDA \cite{nickolls2008scalable} programming---neither the driver nor the application needs to be aware of which host a xPU physically belongs to. The key insight: the xPUs the host sees are not real devices. They are shadow devices injected by the Co-Fabric Switch through automatic enumeration.

\subsubsection{Shadow Device}
\begin{figure}[htbp]
\centering
\includegraphics[width=0.48\textwidth]{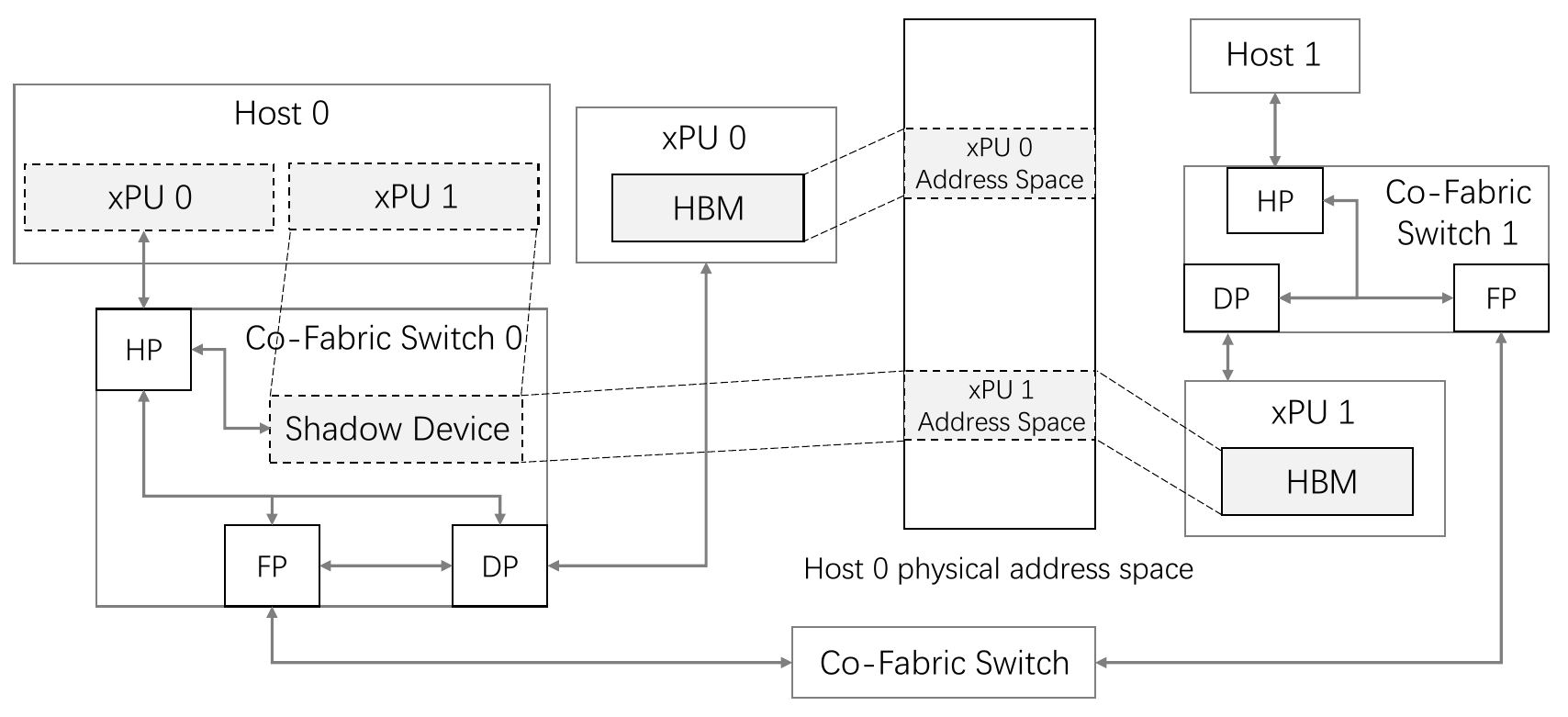}
\caption{Schematic Diagram Of Shadow Device.}
\label{fig:shadow device}
\end{figure}

A shadow device is a virtual endpoint that the Switch presents in the host's device hierarchy, as illustrated in Fig.~\ref{fig:shadow device}. The Switch allocates an HBM address range for it—the range that xPU occupies in the global address space. Any access falling in this range gets routed to the real physical xPU. Every host enumerates the same set of shadow devices from the same group of Switches, and the HBM address ranges are uniformly predefined. So every host sees the exact same address view after enumeration. This is why each OS sees all xPUs as if they were on the same host.

``Every host sees all xPUs" is made possible by the Co‑Fabric Switch's shadow device auto‑enumeration. Within a Co‑Fabric topology, switches may connect to hosts, xPUs, or other switches. During initialization, each Switch enumerates its downstream xPUs and allocates resources on its own. This is shadow device auto‑enumeration. At power‑on, the Switch enumerates downstream xPUs and builds an internal virtual device hierarchy, predefining a consistent HBM address range per xPU for every host's view.

From the xPU side, communication does not change at all. A xPU does not know whether its peer is local or remote, nor does it see the Switch. It just sends Co-Fabric packets following the standard request/completion protocol, with the target address in the HBM range the Switch allocated. The Switch forwards the packet to the real xPU (routing details in the preceding ``Cross-Domain Scaling and ID-oriented Forwarding" section). This is transparent to the xPU—no changes to its software stack or hardware. Co-Fabric is compatible with xPU configuration space models, so existing drivers and runtimes work without adaptation.

At the programming level, the global unified address space takes multi-device collaboration from ``copy data explicitly" to "access by address." Developers don't need to tell local from remote xPUs, write topology-aware code, or set up control objects like Memory Regions (MRs), Queue Pairs (QPs), Completion Queues (CQs), or remote keys (rkeys). The Switch handles all control-path configuration in hardware. Writing distributed xPU code becomes nearly as simple as single-node multi-xPU code.

For access latency, the global address space combined with Co-Fabric's native memory semantics simplifies the cross-domain path. Host read and write requests enter the switching fabric directly as Co-Fabric packets. The Switch does address translation and forwarding—no CPU or driver intervention needed. Addressing and forwarding run in parallel on switch hardware, keeping cross-host xPU access latency in the hundreds of nanoseconds to microseconds.

\subsubsection{Address Space Model}
Co-Fabric uses a flat addressing model where all xPU memory in a superpod maps into a single physical address space, each xPU occupying a non-overlapping range. An address uniquely resolves to a specific xPU and its HBM location. In contrast, traditional distributed systems assign each node its own address space, requiring explicit address mapping for every cross-node access, which adds setup latency and couples code to topology.

The address space is built from uniformly allocated HBM address ranges. During auto‑enumeration, the Switch allocates a fixed‑size HBM range per xPU; all ranges are non‑overlapping. The address space scales gracefully to thousands of xPUs, and range size is configurable per xPU memory capacity. Allocation happens once at init and never changes; the layout is fixed at power‑on. Hosts and xPUs access any xPU by address directly, with no runtime negotiation.

Co-Fabric's unified addressing eliminates the runtime registration overhead of RDMA. RDMA requires developers to register memory regions (MRs), obtain remote keys (rkeys), and include them in every operation. In Co-Fabric, HBM address ranges are predefined at init, so software requires no registration or key lookup at runtime. xPUs read and write by address; the Switch forwards based on predefined mappings. This reduces communication setup latency to near-zero, eliminating the need for runtime negotiation.

\subsection{Low‑Latency Communication \& Link‑Layer Reliability Mechanisms}
\subsubsection{Low Latency}
Co-Fabric enables unified addressing of all xPU memory resources within a superpod, constructing a globally consistent unified address space. Leveraging the high-performance underlying routing mechanisms of the protocol, any xPU can directly access remote data via native local memory instructions. This approach bypasses multi-layer privilege forwarding and address translation procedures, substantially reducing the latency of cross-device data addressing and communication link setup. All memory consistency transactions are completely offloaded to the hardware link layer, which provides fundamental hardware-level low-latency guarantees for cross-node xPU communication.

Based on its native memory semantics, Co-Fabric allows applications to access remote xPU memory directly without extra adaptations for remote procedure calls, intermediate data copying, or interface encapsulation. Developers are freed from manually partitioning data shards and implementing cross-device data migration logic, as remote memory addressing and data transfer are handled directly by hardware and the protocol stack. This eliminates application-layer forwarding overhead and reduces cross-domain accelerator communication latency to the sub-microsecond level. It greatly simplifies the programming model for distributed AI inference and training, lowering the development threshold for applications.

In terms of streamlined protocol design, Co-Fabric abandons the redundant packet encapsulation, multi-stage packet verification, and cumbersome state interaction logic inherent in traditional interconnection protocols. It simplifies packet transmission formats and builds a lightweight semantic layer with native load/store instruction compatibility, further reducing the processing overhead of the communication protocol stack. Single-hop link processing latency is minimized at the fundamental packet transmission stage. Optimized for scale-up superpod, the protocol introduces no additional protocol overhead or latency jitter as the networking scale expands.

\subsubsection{High Reliability}
As the fundamental underlying specification for the superpod interconnect in this work, the Co-Fabric protocol natively integrates a dual reliability assurance mechanism within its link layer, establishing an end‑to‑end robust transmission framework against two primary risk categories—traffic overload across massive node interconnections and physical link bit errors. The first mechanism is a credit‑based flow control scheme, wherein the transmitting and receiving ends periodically exchange link‑status feedback frames to dynamically synchronize the remaining capacity of the receive buffer. A transmitting side is permitted to issue fabric‑layer data units only when the corresponding credit quota at the receiver is sufficient, thereby fundamentally preventing data discards caused by receive‑buffer overflow and eliminating reliability degradation due to congestion‑induced packet loss. The second mechanism is a link‑level automatic retransmission scheme based on positive‑acknowledgment/negative‑acknowledgment (ACK/NAK) responses. The protocol assigns a unique sequence number to each data unit and appends a cyclic redundancy check code; the sender retains unacknowledged copies in a retransmission buffer, while the receiver returns either a positive acknowledgment or a negative retransmission request according to the validation result. Upon detection of bit errors, data‑unit loss, or sequence‑number discontinuities, the link layer autonomously initiates selective retransmission without involving upper‑layer software in error recovery, substantially reducing fault‑repair latency.

The two reliability‑ensuring mechanisms of Co‑Fabric operate in a coordinated manner: credit‑based flow control reduces the probability of triggering retransmission by constraining the injection rate, whereas link‑level retransmission effectively repairs transient errors induced by signal attenuation, crosstalk, or connector contact imperfections. While maintaining low overhead, the Co‑Fabric protocol confines bit‑error recovery latency to the \si{\micro\second} level, and both credit updates and retransmission acknowledgments are embedded within link‑control frames without consuming data‑channel bandwidth. Consequently, under the stringent requirements of high throughput and low latency in superpod environments, the protocol achieves an optimal balance between performance and reliability, thereby establishing a solid data‑transmission foundation for large‑scale high‑availability computing clusters.

\section{Co‑Fabric AI Scaling System}
\label{sec:prototype}
Based on the Co-Fabric architecture, this paper presents the engineering implementation of multiple self-defined Co-Fabric scaling AI server systems, which uniformly integrate xPUs and CPUs into a Co-Fabric interconnect framework featuring high bandwidth, low latency, and a unified memory address space. These systems are built upon a set of supporting hardware components, primarily comprising CPUs and xPUs that are equipped with Co-Fabric-compatible ports, along with switch chips capable of supporting the interconnection of multiple Co-Fabric ports. Leveraging this hardware foundation, this work constructs scaling AI server systems with both southbound-only and hybrid southbound-northbound interconnect architectures, where the northbound direction is defined as the interconnect toward the CPU and the southbound direction as the interconnect toward other components. In this design, the ID routing and address-mapping functions of the Co-Fabric layer are offloaded to Co-Fabric-enabled switches, thereby supporting commercial off-the-shelf computing components. The resulting interconnect fabric exhibits favorable scalability, supporting deployments ranging from dozens to hundreds of cards, with the potential to scale to a thousand-card configuration. The system interconnect topology can be flexibly tailored to diverse application requirements, supporting multiple topologies such as CLOS \cite{al2008scalable}, Rail-only \cite{wang2024rail}, and Full Mesh \cite{hosoki2026optimizing}. These configurations provide xPUs with diversified southbound and northbound Co-Fabric interconnect paths, thereby enabling high-bandwidth, low-latency communication across all xPUs while ensuring cross-domain visibility of xPU resources.

\subsection{Southbound Co-Fabric Scaling System}
\begin{figure}[htbp]
\centering
\subfloat[]{\includegraphics[width=0.34\textwidth]{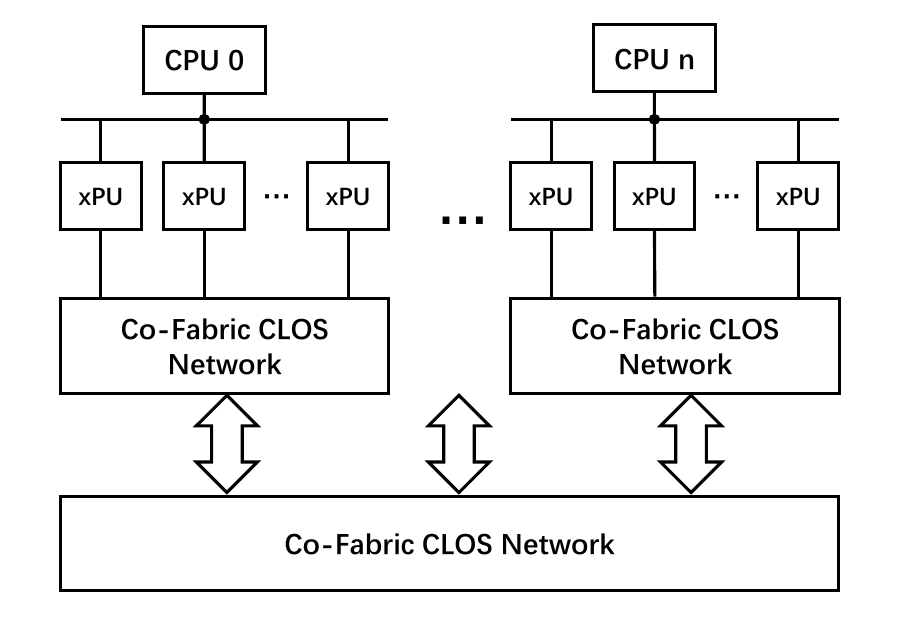}\label{fig:scaling_topology_a}}
\hfill
\subfloat[]{\includegraphics[width=0.34\textwidth]{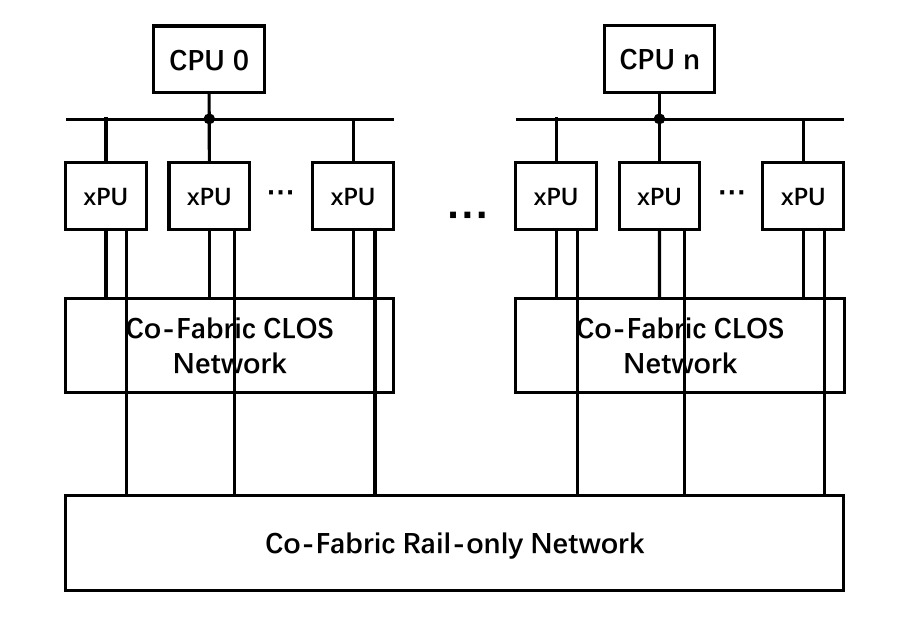}\label{fig:scaling_topology_b}}
\caption{Southbound Co-Fabric scaling system architecture: (a) Two-tier CLOS-based cross-domain symmetric high-bandwidth interconnect; (b) Intra-domain CLOS \& cross-domain Rail-only for asymmetric high-scalability interconnect.}
\label{fig:scaling_topology}
\end{figure}

Employing xPUs that integrate only southbound Co-Fabric ports in conjunction with Co-Fabric switches, this paper constructs a cross-domain southbound scaling system for xPUs, as illustrated in Fig.~\ref{fig:scaling_topology}. Specifically, within a single domain, xPUs are interconnected via a single-tier Co-Fabric switch to form a full-mesh topology among multiple cards, where the number of switches is determined by the number of ports integrated on each xPU, thereby ensuring symmetric high-bandwidth communication among xPUs within the domain. For cross-domain xPUs interconnect, the topology can be flexibly selected according to application scenarios and scaling requirements: (1) As shown in Fig.~\ref{fig:scaling_topology_a}, a two-tier Co-Fabric CLOS network is adopted to maintain symmetric interconnect across domains, guaranteeing high communication bandwidth among xPUs. Based on this approach, we have achieved 32-card scaling interconnect within a single server and cabinet-level deployments exceeding one hundred cards. (2) As depicted in Fig.~\ref{fig:scaling_topology_b}, targeting the typical communication traffic characteristics of large-scale model inference, the interconnect topology is streamlined by employing a Rail-only inter-domain network, which establishes cross-domain connections exclusively between xPUs of the same index, thereby constructing asymmetric communication paths that reduce interconnect cost while expanding the system scale. With this configuration, we have realized scaling to over one hundred cards. The core advantage of the southbound Co-Fabric scaling system lies in its architectural simplicity: the interconnect topology is confined to the xPU side, requiring no CPU involvement in the Co-Fabric data path, which substantially simplifies the complexity of hardware deployment.

\subsection{Hybrid Southbound-Northbound Co-Fabric Scaling System}
\begin{figure}[htbp]
\centering
\includegraphics[width=0.48\textwidth]{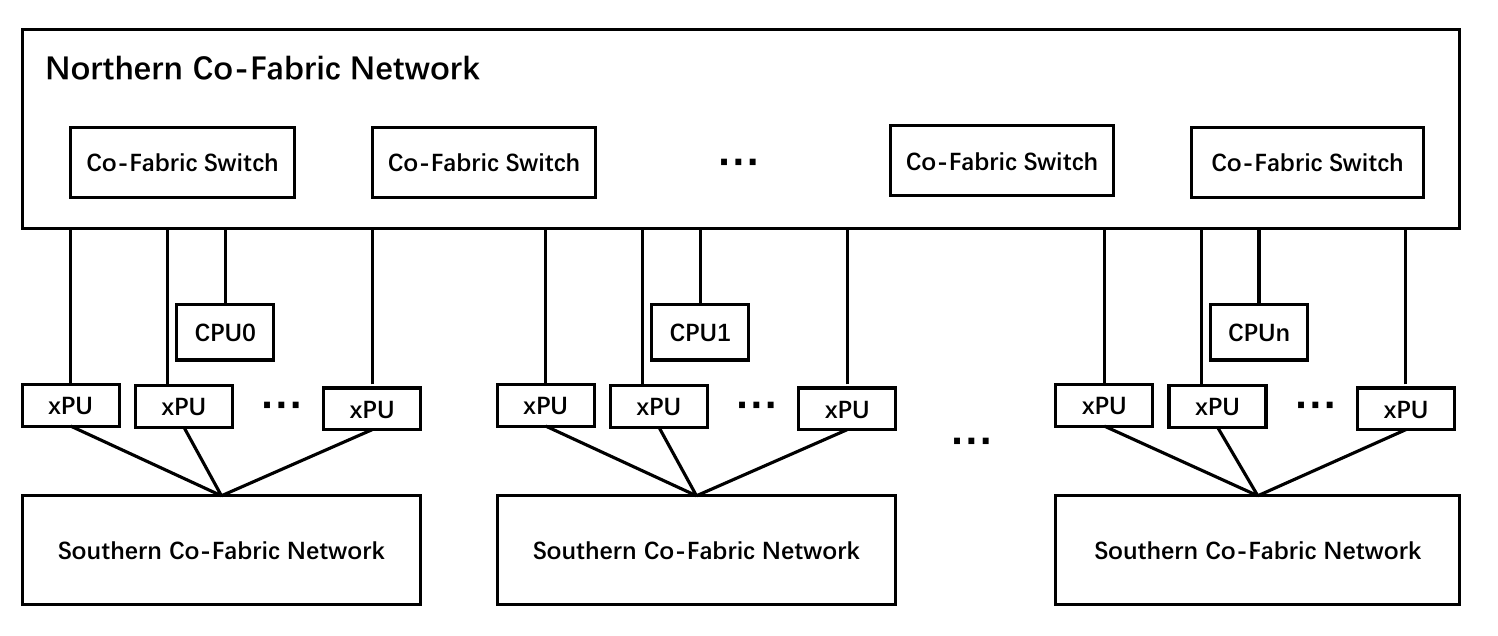}
\caption{Hybrid southbound-northbound Co-Fabric scaling system architecture.}
\label{fig:southbound_northbound_scaling}
\end{figure}

By employing xPUs that integrate both southbound and northbound Co-Fabric ports, CPUs equipped with Co-Fabric ports, and Co-Fabric Switches, this paper constructs a cross-domain hybrid southbound-northbound scaling system, as illustrated in Fig.~\ref{fig:southbound_northbound_scaling}. Specifically, within a single domain, xPUs leverage their southbound Co-Fabric ports to establish a southbound Co-Fabric interconnect network, which can be realized either through Co-Fabric switches or via direct xPU-to-xPU connections in topologies such as Full Mesh. At the cross-domain level, xPUs and CPU connect to the northbound Co-Fabric network—composed of Co-Fabric switches—through both their northbound Co-Fabric ports, thereby enabling cross-domain access. The cross-domain network topology can be flexibly configured according to port count and scaling scale, with options including CLOS, Rail-only, and other variants. Based on this architecture, we have achieved engineering-scale interconnects at the hundred-card level (with detailed system design and test results presented in Section V), with the capability to scale to a thousand-card configuration. Compared with the southbound-only scaling approach, the hybrid interconnect simultaneously utilizes both the southbound and northbound Co-Fabric ports of xPUs, providing a greater number of communication paths among xPUs. This yields substantial advantages in both communication performance and path redundancy, enabling higher bandwidth and lower-latency data transmission. Moreover, with CPUs integrated into the northbound Co-Fabric network, global visibility and efficient sharing of both xPU and CPU resources can be achieved within a unified memory address space.

\section{Evaluation Results and Analysis}
\label{sec:analysis}

Among the two system architectures presented above—namely, the southbound-only and the hybrid southbound-northbound configurations—the hybrid approach more fully exploits the property of the Co-Fabric architecture. In this section, we take the cross-node 64 xPUs hybrid southbound-northbound Co-Fabric scaling system as a concrete example and systematically evaluate its performance advantages from the following dimensions: interconnect link advantages (including link TCO, power comsumption), basic communication performance, system scaling linearity, large-model inference application performance.

\subsection{System Configuration and Comparison Baseline}
\begin{figure}[htbp]
\centering
\subfloat[]{\includegraphics[width=0.16\textwidth]{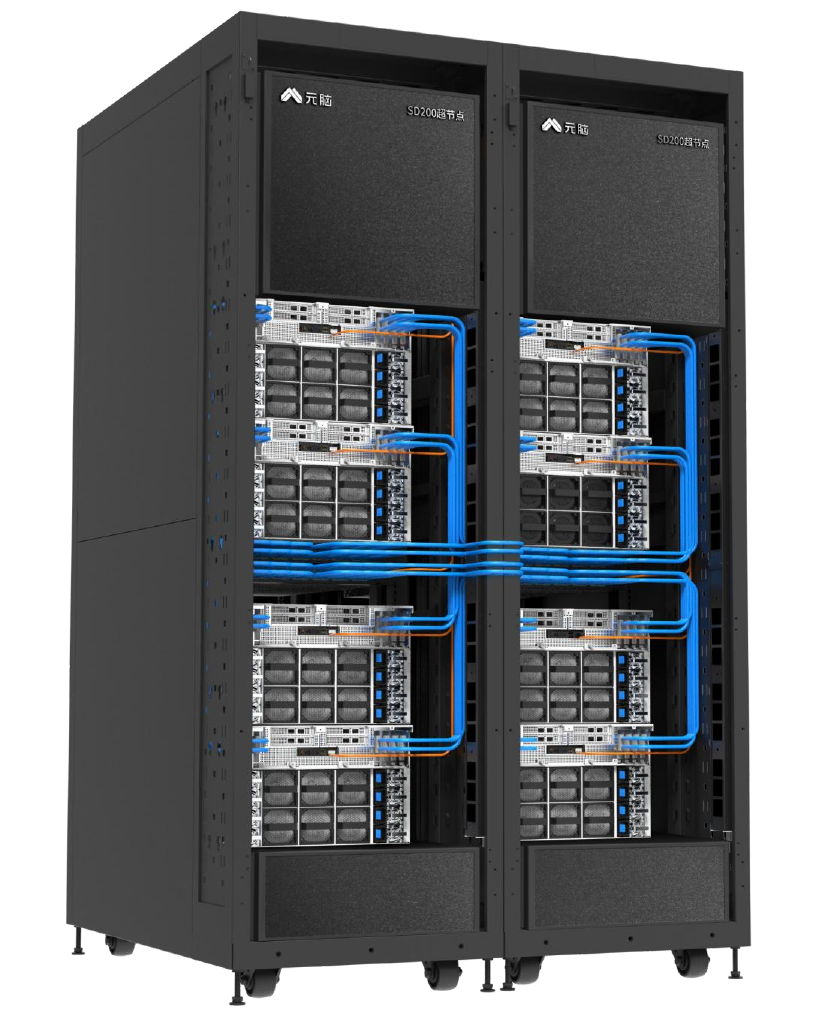}\label{fig:caninet}}
\hfill
\subfloat[]{\includegraphics[width=0.28\textwidth]{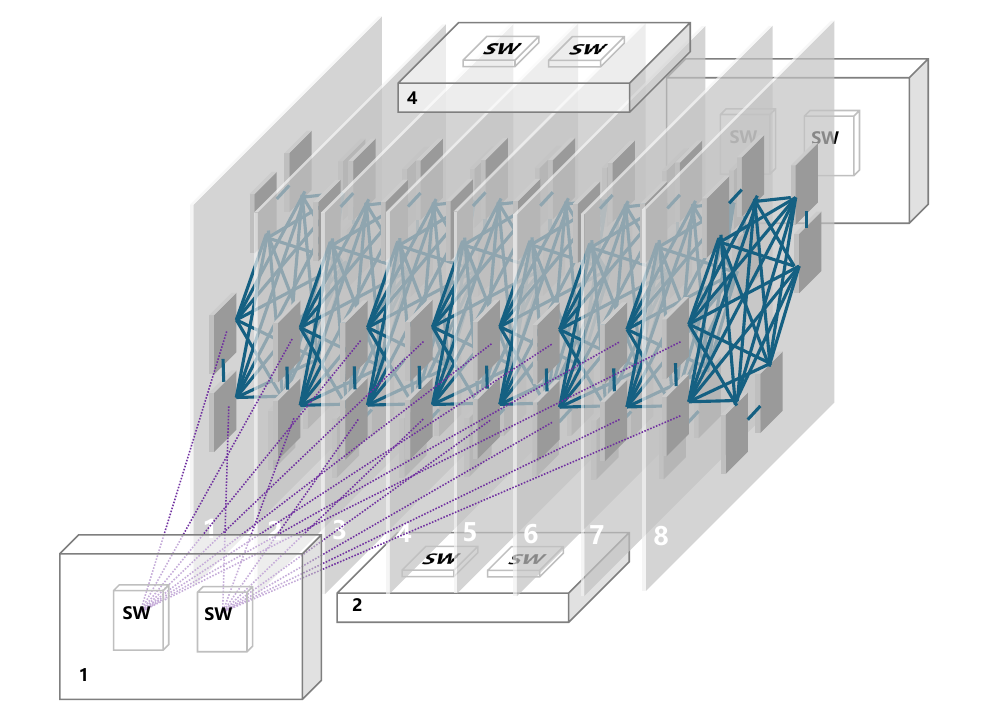}\label{fig:3D_Mesh_topology}}
\caption{The 64 xPUs 3D Mesh Co-Fabric Scaling system and its interconnect topology: (a) Schematic diagram of dual‑cabinet; (b) Schematic of 3D‑Mesh Topology.}
\label{fig:3D_Mesh}
\end{figure}

As shown in Fig.~\ref{fig:caninet}, the 64 xPUs Co-Fabric Scaling system is deployed in a dual-cabinet configuration, with each cabinet housing four 8-xPUs server nodes and two high-performance Co-Fabric switches. The server nodes and switches are physically interconnected via DAC (Direct Attach Copper) cables \cite{kumar2020intra}. At the topology design level, as illustrated in Fig.~\ref{fig:3D_Mesh_topology}, the system adopts a 3D Mesh topology guided by traffic characteristics: within a single node, a southbound Co-Fabric Full Mesh topology is employed to achieve point-to-point direct connections among the eight xPUs; at the cross-node level, given that cross-node communication in large-model inference is typically concentrated among xPUs with the same index, a northbound interconnect based on the Rail-only topology is implemented. This hybrid topology design ensures high efficiency on critical communication paths while effectively controlling interconnect complexity and deployment costs.

To objectively evaluate the performance advantages of the Co-Fabric scaling system, we select a 64 xPUs system based on RoCE interconnect as the comparison baseline. The core difference between the two systems lies in the fact that the RoCE baseline system adopts conventional 8-card OAM servers that do not support Co-Fabric interconnect. Each node provides standard Ethernet ports (rather than Co-Fabric ports) for cross-node scaling through Ethernet NIC cards, and the eight nodes are interconnected via Ethernet switches to construct a 64 xPUs scaling network. A detailed hardware configuration comparison of the two systems is presented in Table~\ref{tab:hw_compare_cofabric_roce}.

\begin{table*}[tbp]
\centering
\caption{Hardware configuration comparison between the 64 xPUs 3D Mesh Co-Fabric Scaling system and the 64 xPUs RoCE scaling system}
\label{tab:hw_compare_cofabric_roce}
\newlength{\colwidth}
\setlength{\colwidth}{\dimexpr(\linewidth - 2.2em)/3\relax}
\begin{tabular}{l p{\colwidth} p{\colwidth}}
\toprule
\textbf{Item} & \textbf{Co-Fabric Scaling System} & \textbf{RoCE Scaling System} \\
\midrule
Scale & 64$\times$xPUs & 64$\times$xPUs \\
\midrule
CPU & 2$\times$CPU/single node, each CPU supports 56 Cores & 2$\times$CPU/single node, each CPU supports 56 Cores \\
xPU & 8$\times$xPUs/single node, each xPU supports 64GB HBM & 8$\times$xPUs/single node, each xPU supports 64GB HBM \\
Memory & 32 $\times$ 64GB DDR5 / single node & 32 $\times$ 64GB DDR5 / single node \\
System Drive & 2$\times$ 960GB SATA SSD / single node & 2$\times$ 960GB SATA SSD / single node \\
Data Drive & 8$\times$ 3.84TB NVMe U.2 / single node & 8$\times$ 3.84TB NVMe U.2 / single node \\
Storage Network & 1$\times$ 400Gbps single‑port Ethernet NIC & 1$\times$ 400Gbps single‑port Ethernet NIC \\
Management Network & 1$\times$ 25Gbps dual‑port Ethernet NIC & 1$\times$ 25Gbps dual‑port Ethernet NIC \\
Single node external interface & 8$\times$ Retimers with 8$\times$Co‑Fabric x16 lane 32Gbps/lane Ports & 8$\times$ Ethernet NICs with 8$\times$ 400Gbps Ports \\
Switches & 4$\times$ Co‑Fabric Switches, each contains 2$\times$ Co‑Fabric Switch chips, with each chip supporting at least 8$\times$ 16‑lane 32Gbps/lane Co‑Fabric ports & 1$\times$ Ethernet Switch, supports at least 64$\times$ 400Gbps ports \\
Interconnection between nodes and switches & 64$\times$ Copper or Optical Cables & 64$\times$ Copper or Optical Cables \\
\bottomrule
\end{tabular}
\end{table*}

\subsection{Analysis of Interconnect Link Advantages}
\begin{figure}[htbp]
\centering
\subfloat[]{\includegraphics[width=0.22\textwidth]{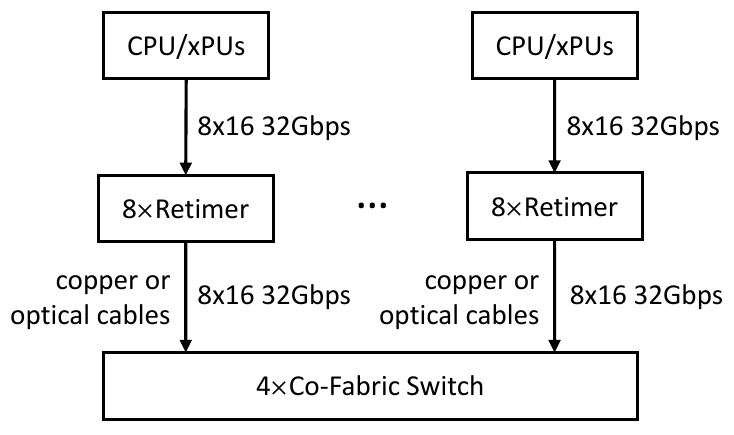}\label{fig:linka}}
\hspace{0.02\textwidth}
\subfloat[]{\includegraphics[width=0.22\textwidth]{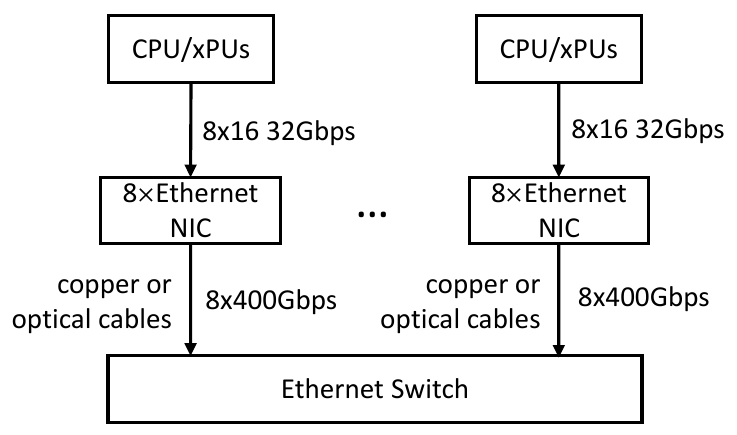}\label{fig:linkb}}
\caption{Interconnection link diagram of the (a) Co-Fabric scaling system and (b) RoCE scaling system.}
\label{fig:link}
\end{figure}

The specific differences in interconnection schemes between the Co-Fabric scaling system and the RoCE scaling system are illustrated in Fig.~\ref{fig:link}. In the Co-Fabric system, the Co-Fabric ports are fanned out via the Retimer cards in the node: the x16 lane, 32 Gbps/lane signals emitted by the CPU/xPUs within a single node, pass through the Co-Fabric ports provided by the eight Retimer cards in the node, and are then connected to the external Co-Fabric switches via copper or optical cables, with no data rate loss across the entire interconnect link. In contrast, in the RoCE system, the data of a single-node CPU/xPUs is routed through the 400 Gbps port of the Ethernet NIC and then connected to the external Ethernet Switch via optical modules and fibers; however, due to the limitations of the Ethernet chip, the signal rate fanned out from the CPU/xPU suffers a certain degree of loss as soon as it is emitted from the node. Such interconnection differences inherently provide the Co-Fabric system with an advantage in bandwidth.

\begin{figure}[htbp]
\centering
\subfloat[]{\includegraphics[width=0.22\textwidth]{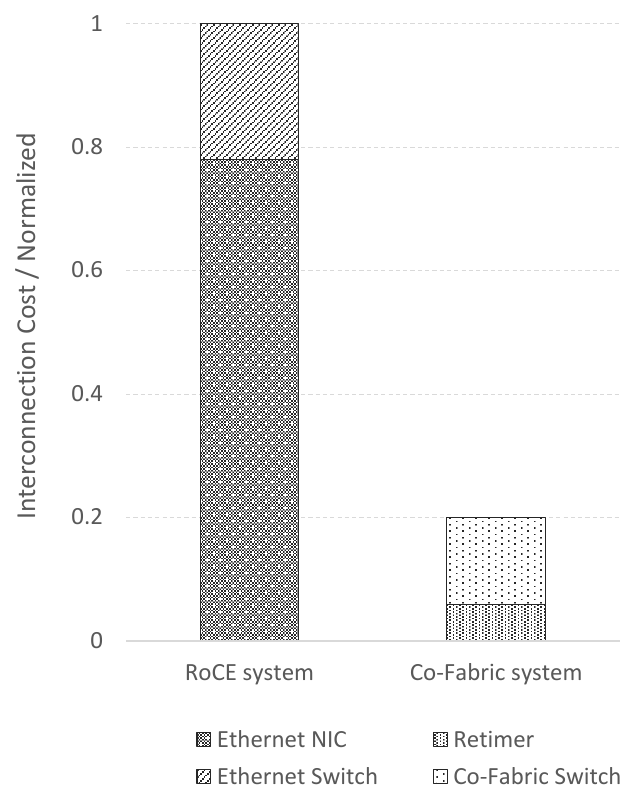}\label{fig:cost}}
\hspace{0.02\textwidth}
\subfloat[]{\includegraphics[width=0.22\textwidth]{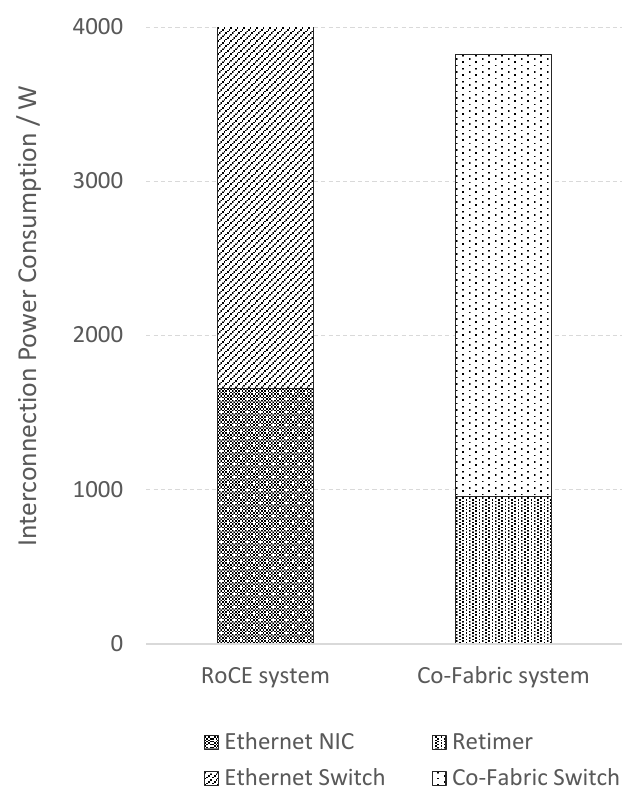}\label{fig:consumption}}
\caption{(a) Comparison of interconnection costs and the corresponding cost breakdown between the Co-Fabric scaling system and the RoCE scaling system; (b) comparison of power consumption and the corresponding power breakdown between the Co-Fabric scaling system and the RoCE scaling system.}
\label{fig:cost_consumption}
\end{figure}

Moreover, additional advantages in TCO and power consumption are achieved on the interconnect link. The Co-Fabric system replaces the high-cost, high-power components  of the RoCE system (Ethernet NIC and Ethernet switch ) with low-cost, low-power alternatives (Retimer and Co-Fabric switches). Fig.~\ref{fig:cost} compares the interconnection costs and the cost breakdown between the RoCE system and the Co-Fabric system, showing that the interconnection cost of the Co-Fabric system is reduced by approximately 80\% relative to the RoCE system. Fig.~\ref{fig:consumption} compares the interconnection power consumption and the corresponding power breakdown of the two systems, showing a reduction in interconnection power consumption of approximately 5\%.

\subsection{Basic Communication Performance Test}

\begin{figure*}[htbp]
\centering
\subfloat[]{\includegraphics[width=0.38\textwidth]{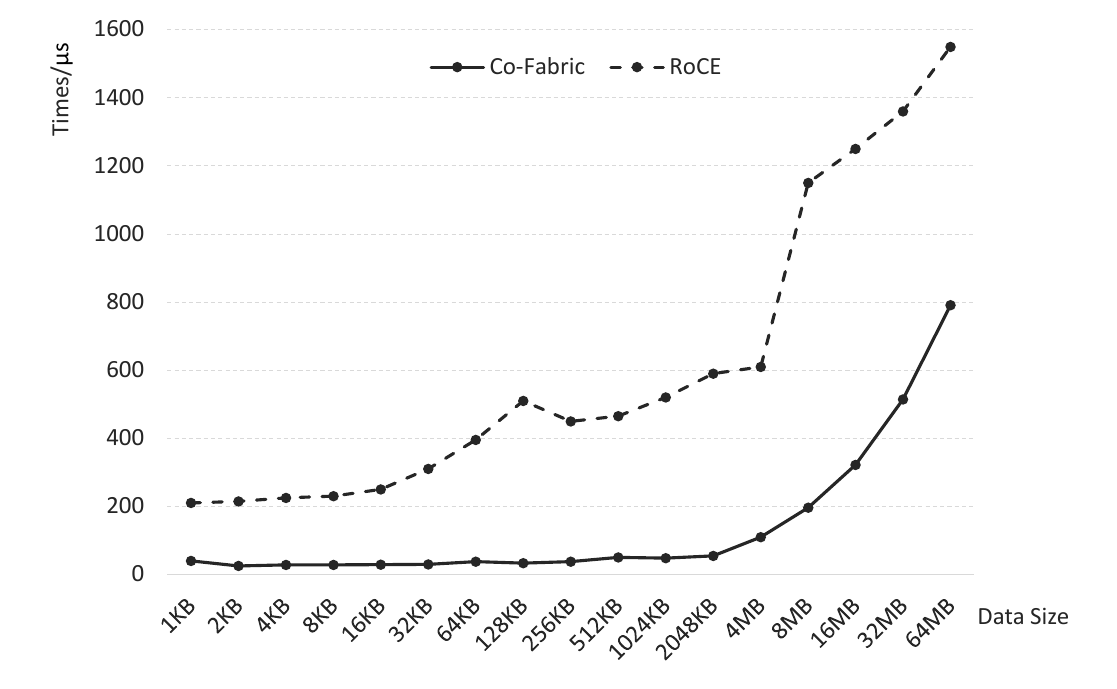}\label{fig:latency_perf}}
\hspace{0.02\textwidth}
\subfloat[]{\includegraphics[width=0.38\textwidth]{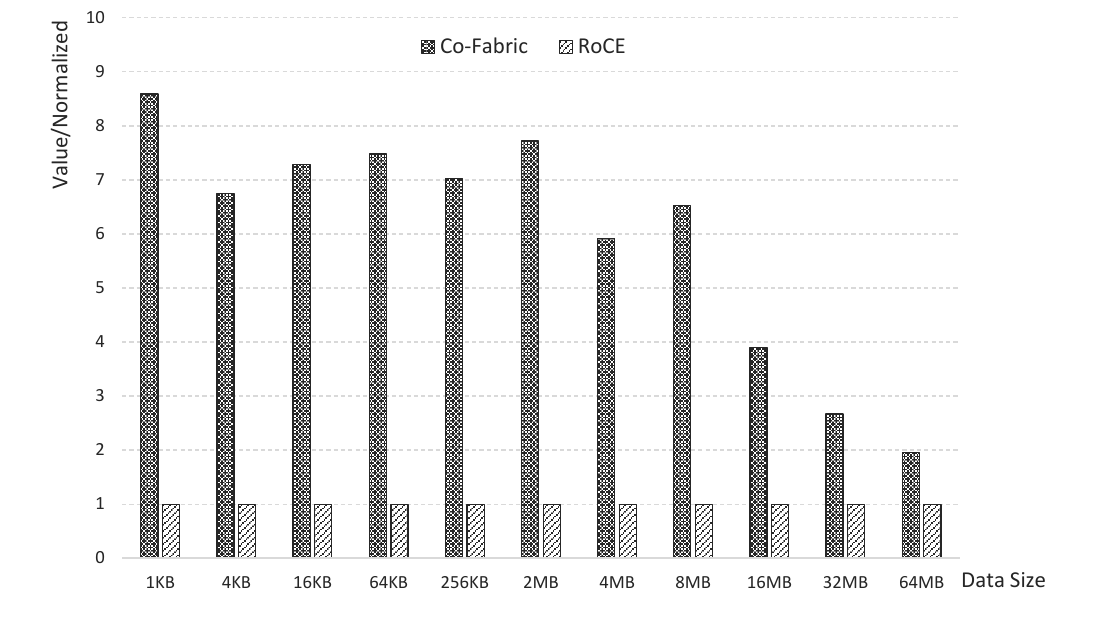}\label{fig:allreduce_perf}}
\caption{Basic communication performance comparison between the 64 xPUs 3D Mesh Co-Fabric scaling system and the RoCE-based scaling system: (a) AllReduce basic communication latency; (b) AllReduce basic communication bandwidth. }
\label{fig:communication_performance}
\end{figure*}

Compared with the traditional RoCE interconnection scheme, the 64 xPUs 3D‑Mesh Co‑Fabric scaling system exhibits prominent advantages of low latency and high bandwidth in terms of basic communication performance. The AllReduce basic communication latency and bandwidth performance tests of the two systems are shown in Fig.~\ref{fig:latency_perf} and Fig.~\ref{fig:allreduce_perf}, respectively. The experimental results show that for small‑packet scenarios below 4MB, the latency of the Co‑Fabric scaling system is only 10\%‑20\% of that of the RoCE scaling system, and its interconnection bandwidth is 5‑15 times that of RoCE. For packet scenarios above 4MB, the latency of the Co‑Fabric scaling system is 15\%‑50\% of that of the RoCE scaling system, and its interconnection bandwidth is 2‑5 times that of RoCE.

\subsection{System Scaling Linearity Test}
This test suite is designed to evaluate the scaling performance of the Co‑Fabric scaling system under representative large‑model workloads at parallel scales of 8, 32, and 64 cards. The test results are summarized in Table~\ref{tab:scaling_linearity}, demonstrating that the Co‑Fabric architecture delivers strong scaling linearity for large‑scale deployments, where communication overhead does not incur notable degradation as the node count increases. For example, in the LoRA fine‑tuning test with DeepSeek V3/R1, the 64‑xPUs Co‑Fabric system achieves a TGS (Tokens per GPU per Second) performance reaching 93.3\% of the 8‑xPUs counterpart. This result shows that performance degradation is bounded within 7\% when scaling the system from 8 to 64 cards.

\begin{table*}[tbp]
\centering
\caption{Scaling linearity test results of the 64 xPUs 3D Mesh Co‑Fabric Scaling system}
\label{tab:scaling_linearity}
\begin{tabular}{l c c c}
\toprule
Model & 8‑Card (Base) & 32‑Card & 64‑Card \\
\midrule
DeepSeek V3/R1, LoRA & Base & 96\% & 93\% \\
QwQ‑32B, Full‑Parameter Mode & Base & 99\% & 98\% \\
Customer Bio‑Genetic Model, Iteration & Base & 99\% & 98\% \\
\bottomrule
\multicolumn{4}{l}{\footnotesize Note: ``Base'' denotes the baseline configuration for scaling‑efficiency calculation.}
\end{tabular}
\end{table*}

\subsection{Large-Model Inference Application Performance Test}
\begin{figure}[htbp]
\centering
\includegraphics[width=0.40\textwidth]{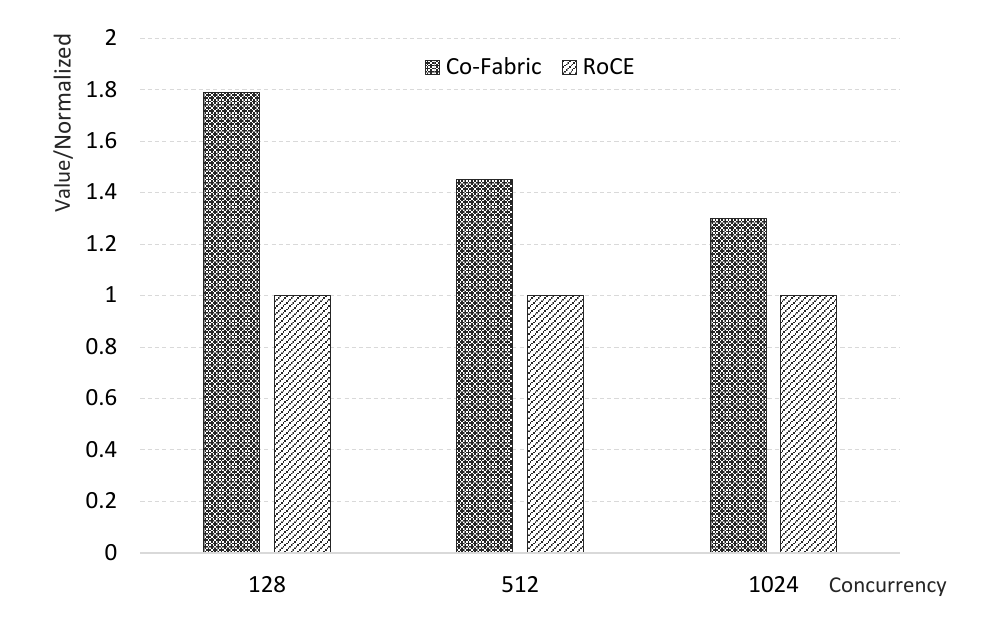}
\caption{DeepSeek R1 inference throughput performance comparison between the 64 xPUs 3D Mesh Co-Fabric scaling system and the RoCE scaling system (Note: DeepSeek-R1 671B inference, SGLang framework, INT8 computing precision adopting MLA, DP4 TP16 and MoE TP parallelism, 1K input/1K output tokens).}
\label{fig:inference_perf}
\end{figure}

To verify the end-to-end benefits of the Co-Fabric architecture in real-world inference workloads, we deploy the DeepSeek R1 \cite{guo2025deepseek} inference model on both systems, with test results presented in Fig.~\ref{fig:inference_perf}. Compared with the RoCE-based Scaling system, the Co-Fabric scaling system achieves an inference performance improvement of approximately 30\% to 80\%. In the 128-concurrency scenario, the inference performance of the Co-Fabric system exceeds that of the RoCE system by a factor of 1.8.

From the perspective of computational logic in large‑model inference, the core process consists of two sequential phases: Prefill and Decode. The communication overheads incurred in both phases collectively determine the end‑to‑end inference computation time. To gain deeper insight into the sources of performance improvement, we compare the overall inference computation time as well as the computation time breakdown across individual phases for the two systems under a fixed workload configuration (batch size = 512, input sequence length = 3072 tokens, output sequence length = 1024 tokens). The results are presented in Fig.~\ref{fig:inference_latency}.

\subsubsection{Prefill phase}
In the Co-Fabric scaling system, the total prefill phase \cite{agrawal2023sarathi} latency is approximately 170 seconds, representing a 26\% reduction compared with 230 seconds in the RoCE system, as shown in Fig.~\ref{fig:prefill_perf}. This gain is primarily attributed to the substantial compression of AllGather communication time, which is reduced from approximately 80 seconds in the RoCE system to merely 20 seconds in the Co-Fabric system, corresponding to a 75\% reduction.

\subsubsection{Decode phase}
\begin{figure}[tbp]
\centering
\subfloat[]{\includegraphics[width=0.22\textwidth]{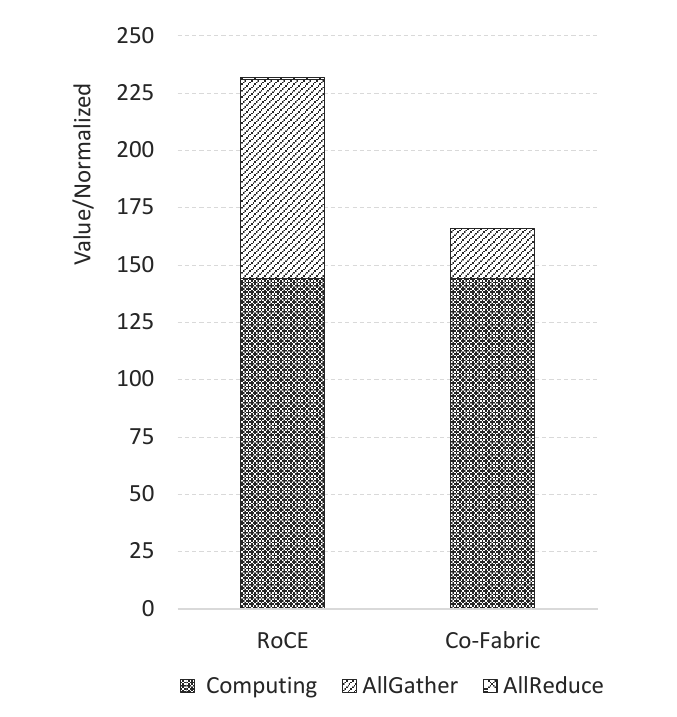}\label{fig:prefill_perf}}
\hspace{0.02\textwidth} 
\subfloat[]{\includegraphics[width=0.21\textwidth]{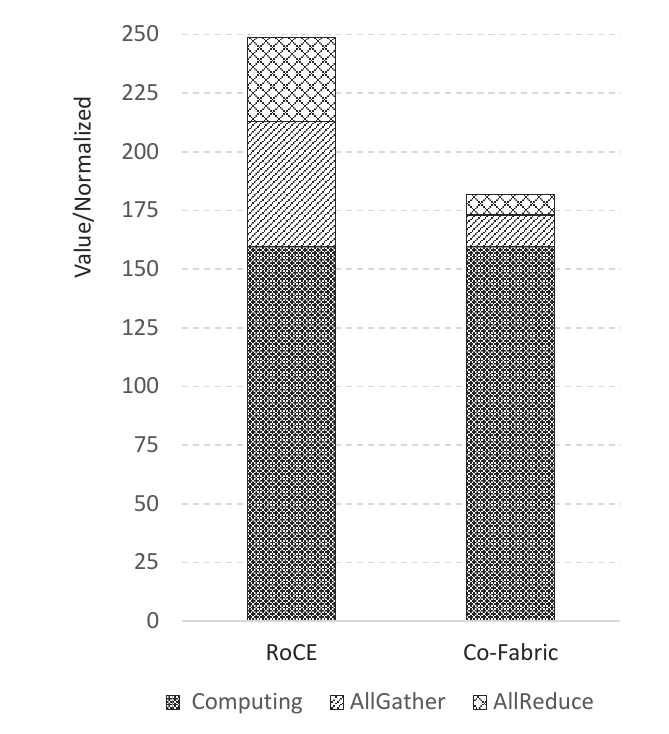}\label{fig:decode_perf}}
\caption{Inference computation time comparison between the 64 xPUs 3D Mesh Co-Fabric scaling system and the RoCE scaling system (unit: second) : (a) prefill phase; (b) decode phase.}
\label{fig:inference_latency}
\end{figure}

In the Co-Fabric scaling system, the total decode phase latency \cite{mcdanel2025pipespec} is approximately 182 seconds, a 25\% reduction compared with 249 seconds in the RoCE system, as shown in Fig.~\ref{fig:decode_perf}. This improvement stems from the significant compression of both AllGather and AllReduce communication operations: in the RoCE system, AllGather and AllReduce consume approximately 53 and 36 seconds, respectively, whereas in the Co-Fabric system, they are reduced to approximately 13 and 9 seconds, respectively, each achieving a 75\% reduction.

\subsection{Summary of Evaluation Results}
Synthesizing the aforementioned experimental results, the following conclusions can be drawn. First, compared with 64 xPUs RoCE scaling system, the Co-Fabric system avoids the signal rate loss introduced by the Ethernet network in its interconnect links, thereby preserving the data bandwidth of the communication links, and achieves 80\% TCO saving and 5\% power consumption savings for intra‑system interconnection though the use of low‑cost, low‑power components. Second, the 64 xPUs Co‑Fabric scaling system significantly outperforms the RoCE scaling scheme in terms of basic communication latency and bandwidth primitives; the latency of the Co‑Fabric scaling system is less than 50\% of that of the RoCE scaling system, and its interconnect bandwidth is 2--5 times that of the RoCE system in large‑packet scenarios (4\,MB-64\,MB). Third, the 64 xPUs Co‑Fabric scaling system demonstrates excellent scaling linearity, maintaining near‑linear performance growth for large model training workloads at the 64 xPUs scale. Finally, at the large‑model inference application level, the Co‑Fabric system achieves a 30\% to 80\% inference performance improvement on the DeepSeek R1 model, with end‑to‑end inference latency effectively reduced. These results collectively validate the effectiveness and superiority of the hybrid southbound‑northbound Co‑Fabric scaling architecture in supporting large‑scale AI workloads.

\section{Conclusion}
\label{sec:conclusion}
This paper presented Co-Fabric, a bus-based interconnect-centric architecture that breaks host-domain boundaries to deliver unified GPU interconnection for scale-up superpods. Co-Fabric makes three contributions. First, a streamlined four-layer protocol stack consolidates compute, storage, networking, and management into a single interconnect domain with nanosecond-scale processing latency and native reliability, decoupling the protocol from physical topology for elastic scaling to thousands of xPUs. Second, cross-domain scaling and P2P communication are realized by embedding port identifiers in the Fabric Layer header, enabling hardware-based routing without host-CPU or software-stack traversal. Third, shadow-device auto-enumeration constructs a global flat address space that lets every host access all xPUs through standard single-node programming models. On a 64-xPU 3D-Mesh system, Co-Fabric's retimer-assisted copper interconnect provides lossless end-to-end transmission and inherent bandwidth advantages over RoCE, together with lower deployment cost and power. It also delivers substantial latency reduction and bandwidth improvement in basic communication, prominent gains for small-size collectives while retaining solid advantages for large-size transfers, near-linear training and fine-tuning speedup, and superlinear inference scaling. DeepSeek R1 inference in particular achieves notable gains with markedly lower end-to-end latency. These results verify that interconnect-centric design provides a viable path to crossing host-domain boundaries and optimizing data movement for large-scale AI systems.

Future work will extend Co-Fabric to larger multi-rack deployments and further offload collective operations onto the fabric.

\section*{Acknowledgment}
The authors are grateful to the reviewers for helpful suggestions on improving the manuscript.

\bibliography{mybib}  

\end{document}